\documentclass[aip,jcp,reprint,nofootinbib]{revtex4-2}
\usepackage{graphicx,amssymb,amsmath}
\usepackage{hyperref}
\usepackage{color}
\usepackage[dvipsnames]{xcolor}
\begin{document}
\title{Ostwald Ripening of Buoyancy-Driven Microbubbles} 
\author{Yuki Uematsu}
\email{uematsu@phys.kyutech.ac.jp}
\affiliation{Department of Physics and Information Technology, Kyushu Institute of Technolohy, Iizuka 820-8502, Japan}
\date{\today}

\begin{abstract}
Microbubble solutions have a wide range of industrial applications, including heat transfer, agriculture, and water treatment.
Therefore, understanding and controlling the size variation of bubbles is critical.
In this study, we develop a theoretical framework for Ostwald ripening in buoyancy-driven microbubbles by introducing a height-dependent size distribution function.
For the first time, we show that the population balance equation in steady state can be interpreted within the Lifshitz–Slezov–Wagner theory when the distribution function is redefined as the density distribution of the buoyancy-induced flux.
Notably, in this form of Ostwald ripening, the distribution function approaches a scaled universal distribution, not over time, but as a function of height.
We analytically derive the scaled universal distribution and show that the fifth power of the mean radius of the bubbles grows linearly with height.
\end{abstract}

\maketitle

Ostwald ripening is a characteristic process of coarsening dynamics in multiparticle systems, where small particles dissolve rapidly, whereas large particles grow slowly due to the Kelvin effect \cite{Ostwald1897,Lifshitz1959,Lifshitz1961,Wagner1961,Burlakov_2006,Lambert_2007,Xu_2017}.
This phenomenon is widely observed in nature, including supersaturated crystal solutions, metal particles in alloys, and oil-in-water emulsions.
Half a century after Ostwald’s seminal work \cite{Ostwald1897}, a quantitative theory describing the asymptotic behavior of coarsening was developed, based on the population-balance equation for the size distribution function \cite{Lifshitz1959,Lifshitz1961,Wagner1961}.
This theoretical framework is now known as the Lifshitz–Slezov–Wagner (LSW) theory. 
The LSW theory demonstrated that the scaled distribution of particle radii converges to a universal shape in the infinite-time limit. 
Furthermore, the cube of the mean radius and the inverse of the number density increase linearly with time.
Recently, various types of Ostwald ripening under nonequilibrium conditions have been studied and developed \cite{Aranson_2002,Zwicker_2015,Tjhung_2018,Saha_2020}.

The dynamics of size variation in multiple bubble systems has been extensively studied \cite{Kocamustafaogullari_1995,Millies_1999,Lehr_2002,Lambert_2007, Inoue_2022} because it has numerous industrial applications, including efficient heat and mass transfer, medical and agricultural applications, and water treatment \cite{Marcelino_2022}.
Theoretical models for these systems are based on the population balance equation.
The main mechanisms of size variation are breakup and coalescence due to the complex flow field \cite{Kocamustafaogullari_1995,Millies_1999,Lehr_2002}.
Although these mechanisms are significant for millimeter-sized bubbles \cite{Kocamustafaogullari_1995,Millies_1999,Lehr_2002}, they are less relevant for microbubbles, which range in size from a few micrometers to tens of micrometers.
In this case, the bubble shape remains spherical during its rise, and the fluid flow around the bubble is laminar and quiescent, resulting in no bubble breakup \cite{clift2005bubbles}. 
Furthermore, when the volume fraction of the bubbles is small, coalescence rarely occurs.
Therefore, Ostwald ripening becomes the main mechanism for size variation in a system of multiple bubbles. 
Indeed, Ostwald ripening of bubbles has been widely observed in aqueous microbubble solutions \cite{Inoue_2022}, dry soap \cite{Lambert_2007}, porous media \cite{Xu_2017,Blunt_2022}, aerated mayonnaise \cite{Dutta_2004}, and magma \cite{Lautze_2010}.
Because the bubbles consist of gas, their mass density is much smaller than that of the surrounding medium.
As a result, the bubbles rise due to buoyancy.  
This effect plays a crucial role in Ostwald ripening, influencing various aspects such as the lateral distribution of bubbles attached to an upper glass surface \cite{Inoue_2022}, gravitational equilibrium \cite{Blunt_2022}, and the height-dependent size distribution function \cite{Dutta_2004}. 
However, no analytical theory has yet been developed for the buoyancy effect on Ostwald ripening.  
Thus, the objective of this Communication is to analytically investigate the buoyancy effect on Ostwald ripening.
We develop a population balance equation for Ostwald ripening in microbubbles rising due to buoyancy by introducing a height-dependent size distribution function.
The steady-state population balance equation is reformulated to be analytically tractable within the LSW theory by using the flux density distribution as a new function. 
Notably, in this Ostwald ripening, the distribution function develops and approaches a scaled universal distribution with increasing height, rather than over time.
The scaled universal distribution was analytically derived, and the fifth power of the mean radius of the bubbles was shown to grow linearly with height. 

\begin{figure}
\includegraphics[width=6cm]{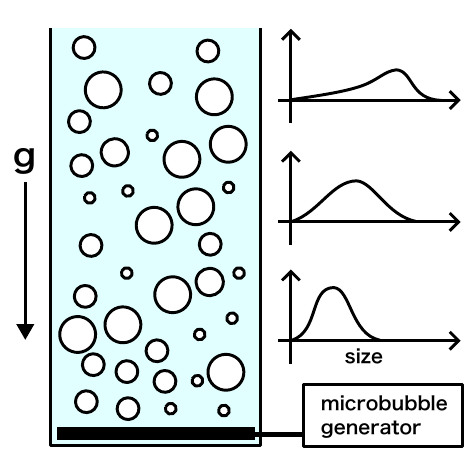}
\caption{
Illustration of microbubbles in a bubble column. 
In steady state, the size distribution function depends on the height. 
}
\label{illust:01}
\end{figure}

We consider a water-filled column, as illustrated in Fig.~\ref{illust:01}, in which bubbles are generated at the bottom of the column, where $z=0$.
Let $F(R,z,t)$ be the size distribution function at height $z$ and time $t$, meaning that $n(z,t)=\int^\infty_0 F(R,z,t)dR$ is the number density of bubbles.
Here, $R$ denotes the radius of the bubbles.
When the volume fraction of bubbles, $\phi(z,t)=\int^\infty_0 \frac{4\pi}{3}R^3 F(R,z,t)dR$, is infinitesimally small, the size dynamics of the bubbles are governed by the local supersaturation $\Delta(z,t)$ and the Laplace pressure \cite{Slezov2005,Inoue_2022}, given by
\begin{equation}
\frac{dR}{dt} = \frac{D}{R}\left(\Delta(z,t) - \frac{\ell}{R}\right),
\label{eq:rdot}
\end{equation}
where $D$ is the diffusion constant of the gas molecules in water.
The term $\ell$ refers to the Kelvin length in bubble systems \cite{Slezov2005,Butt_2009,Inoue_2022}, which is given by $\ell = 2\gamma H/p_0$, where $\gamma$ is the surface tension, $H$ is the dimensionless Henry constant, and $p_0$ is the ambient pressure. 
In Eq.~(\ref{eq:rdot}), the hydrostatic pressure effect, which causes the expansion of the bubbles as they rise, is neglected.
In addition to the size dynamics driven by gas diffusion, the bubbles are also influenced by buoyancy forces.
Assuming that the size of the bubble is sufficiently small and the shape remains spherical, the dynamics of the bubbles in the vertical direction $z$ are given by \cite{clift2005bubbles,Miyazaki_2024}
\begin{equation}
\frac{dz}{dt} = aR^2,
\label{eq:zdot}
\end{equation}
where $a=2\rho g/9\eta$ is the buoyant coefficient, $\rho$ is the mass density of water, $g$ is the acceleration due to gravity, and $\eta$ is the viscosity. 
Here, the mass density of the gas phase is neglected.
At the bubble interface, a no-slip condition is assumed, meaning the velocity of the water at the surface of the bubble is zero.

When bubbles are continuously generated at the bottom of the column, the size distribution function follows a population balance equation given by  
\begin{equation}
\frac{\partial F}{\partial t} = -\frac{\partial}{\partial R}\left[F(R,z,t)\frac{dR}{dt}\right] -\frac{\partial}{\partial z} \left[F(R,z,t)\frac{dz}{dt}\right].
\label{eq:continuity}
\end{equation}
The sum of local supersaturation and volume fraction is governed by the following conservation law,
\begin{equation}
\frac{\partial}{\partial t}\left(\Delta+\phi \right) = -\frac{\partial}{\partial z}\left[\int^\infty_0 \frac{4\pi}{3} R^3 F(R,z,t) \frac{dz}{dt} dR\right].
\label{eq:supersaturation}
\end{equation}
In the steady state, Eq.~(\ref{eq:continuity}) simplifies to
\begin{equation}
\frac{\partial}{\partial z}\left[F(R,z)\frac{dz}{dt}\right] = -\frac{\partial}{\partial R}\left[F(R,z)\frac{dR}{dt}\right],\label{eq:steadycontinuity}
\end{equation}
where $F(R,z)$ no longer depends on the time. 
Equation~(\ref{eq:steadycontinuity}) cannot be analyzed using the LSW theory because the term $dz/dt$ disturbs the form of the continuity equation for the size distribution function $F(R,z)$.

To make Eq.~(\ref{eq:steadycontinuity}) analytically tractable within the LSW theory, we introduce the density distribution of buoyancy-induced flux at $z$ as
\begin{equation}
f(R,z) = F(R,z)\frac{dz}{dt} = aF(R,z)R^2.
\end{equation}
As a result, Eq.~(\ref{eq:steadycontinuity}) can be interpreted as the continuity equation for $f(R,z)$ with respect to the effective time $z$, which is given by
\begin{equation}
\frac{\partial f}{\partial z} = -\frac{\partial}{\partial R}\left[f(R,z) \frac{dR}{dz}\right],\label{eq:lsw1}
\end{equation}
where $dR/dz$ represents the effective flux of $f(R,z)$, and is given by
\begin{equation}
\frac{dR}{dz}=\frac{1}{aR^2}\frac{dR}{dt}=\frac{D\ell/a}{R^3}\left(\frac{1}{R_\mathrm{c}(z)} - \frac{1}{R}\right),\label{eq:lsw2}
\end{equation}
where $R_\mathrm{c}(z) = \ell/\Delta(z)$.
The critical radius $R_\mathrm{c}(z)$ is determined by Eq.~(\ref{eq:supersaturation}) in the steady state, as described by
\begin{equation}
\int^\infty_0 \frac{4\pi}{3} R^3 f(R,z) dR = \textrm{const. with } z.
\label{eq:J0}
\end{equation}
The critical radius, subject to the constraint given in Eq.~(\ref{eq:J0}), can be calculated by substituting Eqs.~(\ref{eq:lsw1}) and (\ref{eq:lsw2}) into Eq.~(\ref{eq:J0}), which results in 
\begin{equation}
R_\mathrm{c} = \frac{\int^\infty_0 R\,F(R,z)dR}{\int^\infty_0 F(R,z) dR} = \langle R \rangle,\label{eq:Rcmean}
\end{equation}
where $\langle \cdots \rangle$ denotes the average with the size distribution function $F(R,z)$.
Thus, the critical radius is equivalent to the mean radius, as is the case in diffusion-limited Ostwald ripening \cite{Lifshitz1959,Lifshitz1961,Wagner1961}. 

A novel finding here is that Eqs.~(\ref{eq:lsw1}) to (\ref{eq:J0}) are identical to the equations of LSW theory with $n=5$ \cite{Lifshitz1959,Lifshitz1961,Wagner1961,Giron1998,Meerson_1999}, which are typically expressed as
\begin{equation}
\frac{\partial f}{\partial t} = -\frac{\partial}{\partial R}\left[\frac{K_n}{R^{n-2}}\left(\frac{1}{R_\mathrm{c}(t)}-\frac{1}{R}\right)f(R,t)\right],
\end{equation}
with the constraint $\int^\infty_0 R^3 f(R,t) dR = \textrm{const.}$, 
where $K_n$ is the kinetic coefficient, and $n$ is an integer representing the type of mass transfer. 
The case $n=2$ corresponds to reaction on the interphase surface (interface-limited), whereas $n=3$ corresponds to volume diffusion (diffusion-limited).
In metallurgy, the cases $n=4$ and $n=5$ correspond to grain-boundary diffusion and diffusion along a dislocation network, respectively \cite{Slezov_1978,Slezov_1987,Alexandrov_2017}.
However, the higher-order cases ($n=4$ and $5$) were purely mathematical models and, to the best of our knowledge, have not been discussed in practical contexts.
Although buoyancy-driven microbubbles are coarsened by volume diffusion of gas molecules, the $R^2$-dependence of the migration velocity in Eq.~(\ref{eq:zdot}) leads to Ostwald ripening with $n=5$.

\begin{figure}
\center
\includegraphics[width=7cm]{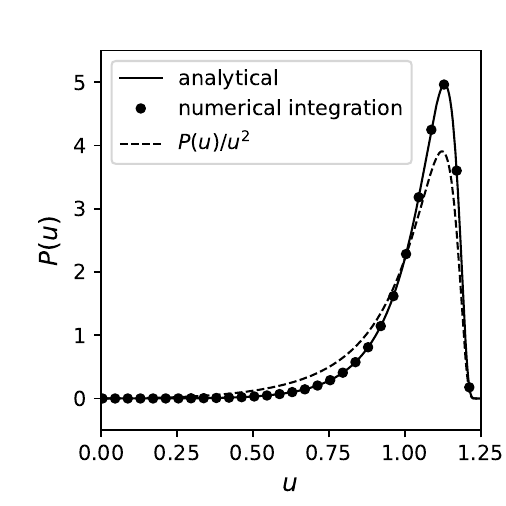}
\caption{The universal distribution $P(u)$. 
The solid line represents the analytical result obtained from Eq.~(\ref{eq:P5}), while the points corresponds to the numerical integration of Eq.~(\ref{eq:Psol}).
The dashed line depicts the universal distribution for the size population distribution $F(R,z)$. 
}
\label{fig:01}
\end{figure}

The LSW theory demonstrates the existence of a scaled universal distribution for the flux density distribution, $f(R,z)$ \cite{Slezov_1978,Slezov_1987,Alexandrov_2017}. 
Below, we derive the universal distribution and discuss its scaling behavior. 
First, the lengths, $R$, $z$, and $R_\mathrm{c}(z)$, are nondimensionalized using the length scale $(D\ell/a)^{1/4}$ as
$r=R/(D\ell/a)^{1/4}$, $\tau=z/(D\ell/a)^{1/4}$, and $r_\mathrm{c}=R_\mathrm{c}/(D\ell/a)^{1/4}$.
The flux density distribution in dimensionless space, $\varphi(r,\tau)$, is defined such that $\int^\infty_0 r^3 \varphi(r,\tau) dr = 1$. 
The LSW theory with $n=5$ \cite{Giron1998,Meerson_1999} predicts the existence of the asymptotic solutions given by 
\begin{eqnarray}
r_\mathrm{c}(\tau) & = & r_*\left(\tau+\tau_0\right)^{1/5},\label{eq:scaling1}\\
\varphi(r,\tau) & = & A r_\mathrm{c}(\tau)^{-4} P(r/r_\mathrm{c}(\tau)),\label{eq:scaling2}
\end{eqnarray}
where $r_*$, $\tau_0$, and $A$ are constants. 
The scaled size distribution function $P(u)$ is defined to satisfy $\int^\infty_0 P(u) du = 1$, where $u=r/r_\mathrm{c}$.
To derive a differential equation for $P(u)$, Eq.~(\ref{eq:scaling2}) is plugged into Eq.~(\ref{eq:lsw1}) to obtain
\begin{equation}
P(u) = \frac{d}{du}\left[G(u;r_*)P(u)\right], \label{eq:steady3}
\end{equation}
where
\begin{equation}
G(u;r_*) = \frac{1}{3}\left(-u+5r_*^{-5}\frac{u-1}{u^4}\right).
\end{equation}
The solution of Eq.~(\ref{eq:steady3}) is 
\begin{equation}
P(u) = \frac{B}{G(u;r_*)}\mathrm{exp}\left(\int^u \frac{du'}{G(u';r_*)}\right),\label{eq:Psol}
\end{equation}
where $B$ is determined by the condition $\int^\infty_0 P(u) du = 1$.
The LSW theory shows that, in the limit as $\tau\to\infty$, nearly any distribution converges to a universal distribution \cite{Giron1998,Meerson_1999}, given by $\varphi(r,\tau) \to Ar_\mathrm{c}(\tau)^{-4}P(r/r_\mathrm{c}(\tau))$, where $r_* = (4/5)^{4/5}$ is determined by the condition that $u^4G(u;r_*)=0$ has multiple roots for $u>0$.
The analytical expression for $P(u)$ is 
\begin{equation}
P(u) = \frac{Cu^4\mathrm{e}^{-\frac{3/2}{5-4u}}\prod_{i=1,2,3}(u-\omega_i)^{\beta_i}}{(5-4u)^{29/10}(16u^3+40u^2+75u+125)},\label{eq:P5}
\end{equation}
for $0\le u\le 5/4$ and $P(u)=0$ for $u>5/4$.
$\omega_i$ ($i=1,2,3$) are the roots of a cubic equation, $16\omega^3+40\omega^2+75\omega+125=0$. 
The exponents $\beta_i$ are given by
\begin{equation}
\beta_i = -\frac{3\cdot 2^8 \omega_i^4}{(5-4\omega_i)^2(48\omega_i^2+80\omega_i+75)},\label{eq:beta}
\end{equation}
and $C = 2^8\cdot 3\cdot 5^{9/10}\cdot\mathrm{e}^{3/10}\prod_i(-\omega_i)^{-\beta_i}= 15545.4$ is the normalization constant.
This analytical expression, given by Eq.~(\ref{eq:P5}), differs from that reported earlier \cite{Alexandrov_2017}.
The derivation and numerical check of eq.~\ref{eq:P5} is explained as below. 
Using Heaviside's expansion theorem, a rational function $1/G(u;(4/5)^{4/5})$ can be expanded as 
\begin{equation}
\frac{1}{G(u;(4/5)^{4/5})} = \frac{\alpha_1}{u-5/4}+\frac{\alpha_2}{(u-5/4)^2}+\sum_{i=1,2,3}\frac{\beta_i}{u-\omega_i},\label{eq:Heaviside}
\end{equation}
where $\alpha_1 = -9/10$, $\alpha_2=-3/8$, and $\beta_i$ are given in eq.~\ref{eq:beta}. 
Plugging eq.~\ref{eq:Heaviside} into eq.~\ref{eq:Psol} gives eq.~\ref{eq:P5}.
Fig.~\ref{fig:01} shows the universal distribution $P(u)$.
The solid line was analytically obtained by Eq.~(\ref{eq:P5}), whereas the points were obtained by numerical integration of Eq.~(\ref{eq:Psol}).
Their agreement demonstrates the correctness of the analytical expression of $P(u)$, Eq.~(\ref{eq:P5}).
The universal distribution for $f(R,z)$ implies a corresponding universal distribution for the size distribution function $F(R,z)$, given by $P(u)/u^2$ and represented by the dashed line in Fig.~\ref{fig:01}.

As a result of the scaling behavior of the distribution function, the fifth power of the mean radius grows linearly with height, as given by 
\begin{equation}
\langle R \rangle ^5  =\left(\frac{4}{5}\right)^4 \frac{D\ell}{a}(z+z_0), 
\label{eq:fifthpower}
\end{equation}
where $z_0 = (D\ell/a)^{1/4}\tau_0$. 
This behavior is similar to the diffusion-limited Ostwald ripening ($n=3$), in which the cubic power of the mean radius grows linearly with time \cite{Lifshitz1959,Lifshitz1961,Wagner1961}. 
The number of bubbles as a function of height is inversely proportional to the height, as given by $n(z)\propto (z+z_0)^{-1}$.
The volume (void) fraction as a function of height follows an exponent of $-2/5$, as given by $\phi(z) \propto (z+z_0)^{-2/5}$.
The literature values for air/water systems at $298\,$K are listed in Table~\ref{tab:01}.
Using the standard pressure $p_0=0.1\,$MPa and gravitational acceleration $g=9.8\,$m/s$^2$, we obtained the slope in Eq.~(\ref{eq:fifthpower}) is $d\langle R\rangle^5/dz = 25^5 \,\mu \mathrm{m}^5/\mathrm{m}$, indicating that coarsening is most significant for micrometer-sized bubbles. 

\begin{table}
\caption{The literature values for the surface tension $\gamma$, dimensionless Henry constant $H$, the mass density of water $\rho$, the viscosity of water $\eta$, the diffusion constant of gas in water $D$. 
The values for air/water systems at 298$\,$K are taken from Ref.~\citenum{Inoue_2022}.
The slopes of Eq.~(\ref{eq:fifthpower}) are also listed.
}
\label{tab:01}

\vspace{3mm}

\begin{tabular}{cc}
 & air/water at 298$\,$K \\\hline
$\gamma\,$(mN/m)& 72 \\
$H$ & $1.9\times 10^{-2}$ \\
$\rho\,$(kg/m$^3$) & 997 \\
$\eta\,$(mPa$\cdot$s) & 0.89 \\
$D\,$(m$^2$/s) & $2.1\times 10^{-9}$ \\
$d\langle R\rangle^5/dz\,$($\mu$m$^5$/m) & $25^5$ \\\hline 
\end{tabular}
\end{table}

In the end, we make a remark on the application of the theory to nanobubbles, which are defined as having a size less than $1\,\mu$m.
The long-term stability of bulk nanobubbles has been debated for years \cite{Epstein1950,Jadhav_2020,Tan_2020,Jaramillo_Granada_2022,Miyazaki_2024}. 
Considering that nanobubbles are generated from the shrinkage of a swarm of microbubbles in the column, Eqs.~(\ref{eq:rdot}) and (\ref{eq:lsw2}) determine the lifetime and traveled distance of a single nanobubble. 
When the initial bubble radius is $R_0=1\,\mu$m ($\ll R_\mathrm{c}$), we obtain $6\,$ms \cite{Epstein1950} and $10\,$nm, implying that the nanobubbles disappear quickly before they rise.
On the other hand, when the nanobubbles are generated as a swarm of monodisperse nanobubbles ($R_0\approx R_\mathrm{c}$), the nanobubble can be stabilized for a long time and over significant distances due to Ostwald ripening.  
However, the experimentally observed size distribution functions of nanobubbles are typically broad \cite{Jadhav_2020}, and this scenario of long-term stability for bulk nanobubbles does not fully explain what actually occurs.   

In summary, we develop an analytical theory of Ostwald ripening in microbubbles rising due to buoyancy by introducing a height-dependent size distribution function.
Starting with the population balance equation, we demonstrate, for the first time, that this equation in the steady state can be interpreted within the framework of LSW theory when the distribution function is redefined as the density distribution of bubble flux.
Notably, in this form of Ostwald ripening, the distribution function approaches a scaled universal distribution as a function of height rather than time.
The scaled universal distribution was analytically derived, and the fifth power of the mean radius of the bubbles grows linearly with height.
We emphasize that, to the best of our knowledge, this is the first study to explore Ostwald ripening along the vertical dimension. 
The formalism demonstrated here is not limited to bubble systems.
It is also applicable to any system where the particles and the surrounding medium have different mass densities, such as emulsions and aerosols where sedimentation can act as the driving force.
Furthermore, to generalize our results, if the particle velocity follows $aR^m$, where $m \neq 2$, the system will exhibit Ostwald ripening with $n=m+3$. 
This suggests a strong possibility of observing higher-order Ostwald ripening, particularly for $n=4$ or $n \ge 6$. 
We hope that these analytical results inspire further experimental observations and verifications of the size distribution function and scaling behavior in driven particle systems, such as electrophoretic or other active particles.

\begin{acknowledgments}
The author thanks financial support from JST PRESTO (Grant No. JPMJPR21O2).
\end{acknowledgments}

\section*{DATA AVAILABILITY}
The data that support the findings of this study are available from the corresponding author upon reasonable request.

\bibliography{ostwald}
\end{document}